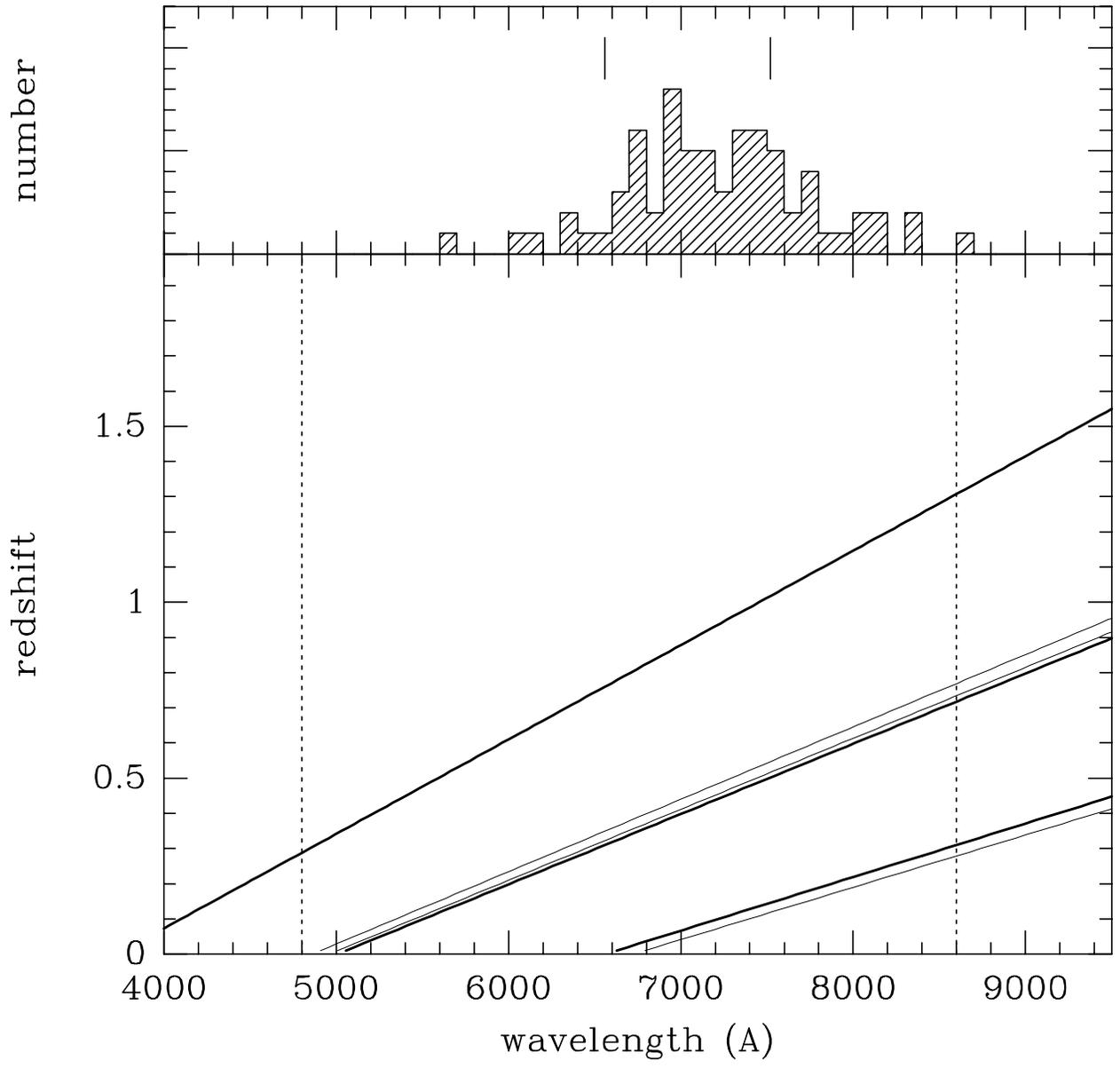

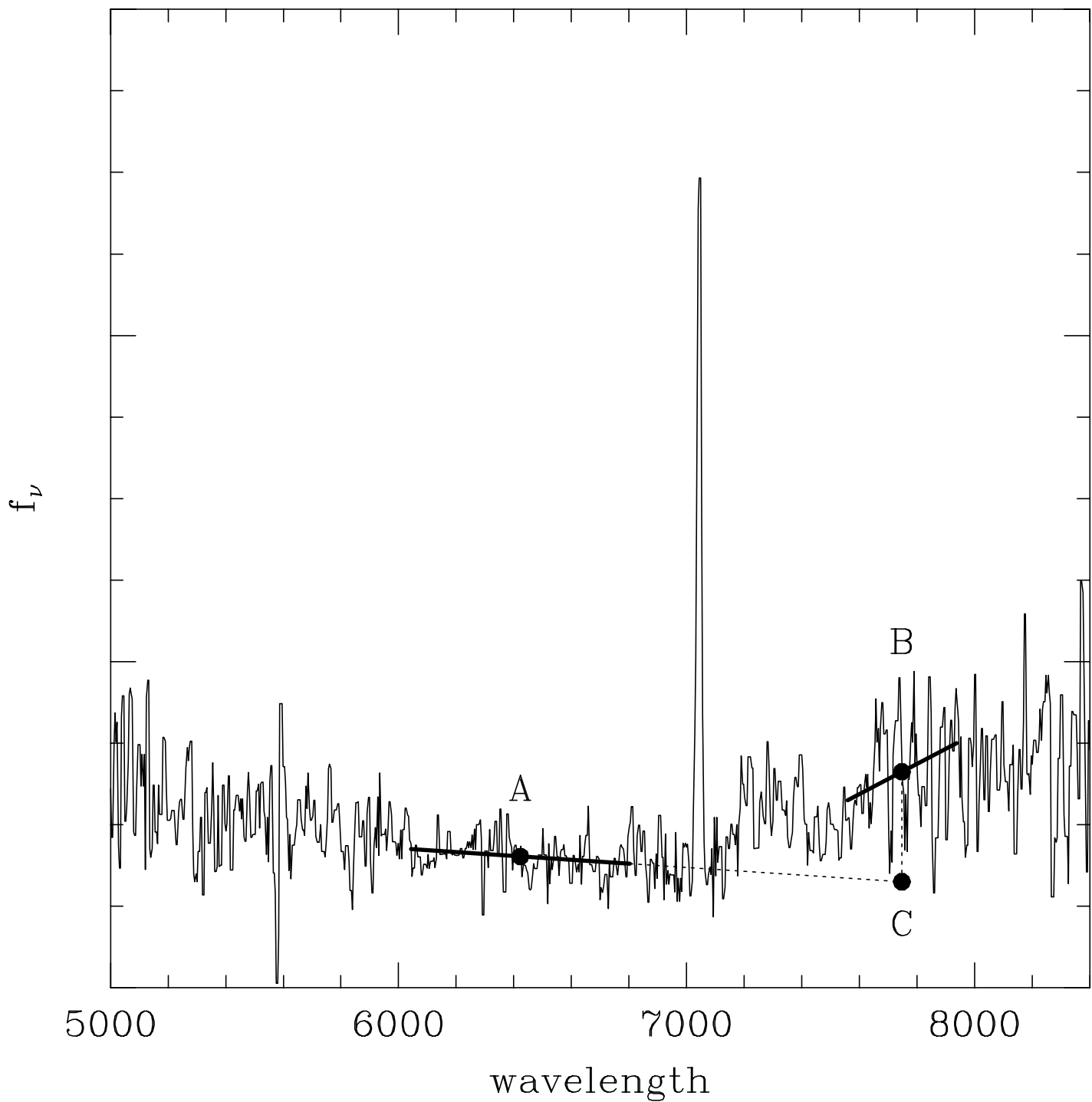

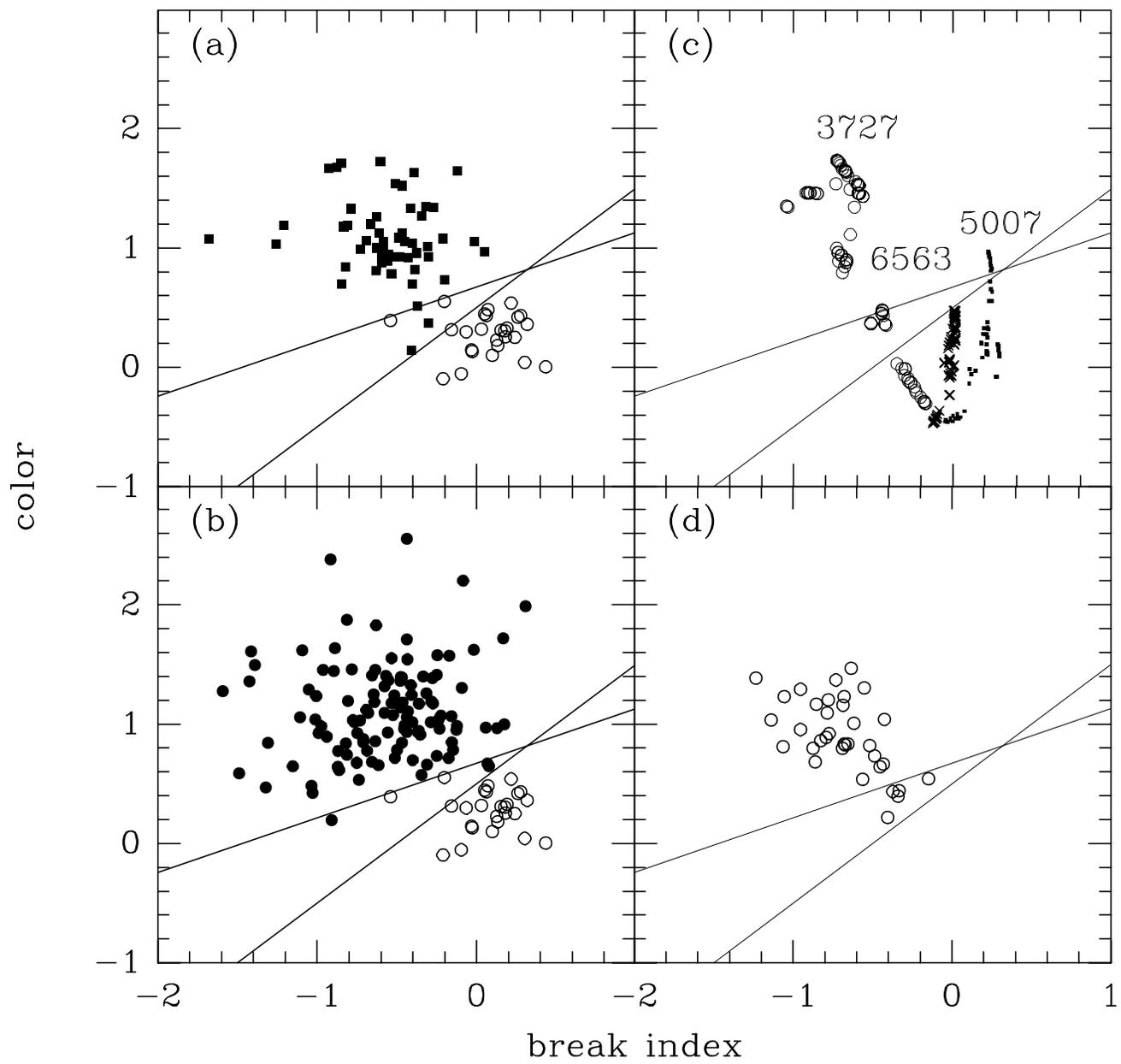

# The Canada-France Redshift Survey III: "single emission line" objects, analysis of repeat observations and spectroscopic identifications in the 1415+52 and 2215+00 fields

S. J. Lilly[1]

Department. of Astronomy, University of Toronto, Toronto, Ontario M5S 1A7, Canada

F. Hammer[1], O. Le Fèvre[1]

DAEC, Observatoire de Paris-Meudon, 92195 Meudon, France

David Crampton[1]

Dominion Astrophysical Observatory, Victoria, British Columbia, Canada

## ABSTRACT

This paper is one of a series describing the Canada-France Redshift Survey (CFRS). It is shown how the shape of the continuum around the emission line can be used to distinguish between [OII] 3727 at z > 0.76 and H$\alpha$ at low redshift in spectra for which only a single isolated emission line is visible. Based on this, [OII] 3727 is most likely to be the emission line in most of the single emission line galaxies in the CFRS. The statistics of the repeated observations are analyzed to derive an empirical calibration of the reliability of the spectroscopic identifications in the CFRS, to determine how often additional observations could lead to the identification of an initially unidentified object and to provide an estimate of the internal velocity accuracy. Finally, the results of spectroscopic observations of 413 objects in the 1415+52 and 2215+00 CFRS survey fields are presented.

*Subject headings:*  galaxies: evolution — galaxies: luminosity function

## 1. Introduction

In this paper, one of a series describing the Canada France Redshift Survey (CFRS), the question of whether the shape of the continuum in the neighborhood of an emission line can be used to identify the emission line is examined. This is of particular importance in view of the significant fraction of objects in this and other faint galaxy redshift surveys for which only a isolated emission line is seen. For emission lines at $\lambda > 6560$ Å there is a potentially ambiguity between [OII] 3727 at z > 0.76 and H$\alpha$ at low redshift.

Many objects in the CFRS were observed on more than one occasion either by design or accident. Each of these multiple observations was initially reduced and identified without reference to any earlier spectra, after which they were co-added (if this was advantageous). The statistics of these multiple observations are examined to determine the reliability of the spectroscopic identifications in the different confidence classes described by Le Fèvre et al (1995a, CFRS II). These repeat observations can also be used to infer the nature of some of the sources that remain unidentified in the final spectroscopic sample, and to provide an estimate of the internal accuracies of the redshift measurements.

Finally, spectroscopic data is presented for a sample of 437 faint stars and galaxies with $17.5 \leq I_{AB} \leq 22.5$ in the 1415+52 and 2215+00 survey fields of the CFRS.

## 2. Galaxies with single isolated emission lines

A familiar problem for those who work in this field is how to treat objects in which the only sharp feature is a single isolated emission line superposed on a continuum. In the absence of any other identifiable absorption or emission lines, can such an emission line be reliably identified?

The CFRS sample of 1010 spectra contains 71 objects in which only a single emission line was seen. Figure 1a shows the distribution of wavelengths of the emission line in these objects. In the majority of objects the emission line was seen towards the red end of the spectrum. While most people would accept the hypothesis that these single isolated emission lines are produced by [OII] 3727, other possibilities exist and, in order to treat these objects as rigorously as possible, all such spectra were put to one side for the systematic analysis described below.

Figure 1b shows the wavelengths of the strongest emission lines usually seen in galactic spectra as a function of redshift within the spectral range of the CFRS spectra. A single isolated emission line at $\lambda > 4800$ Å is unlikely to be produced by one of H$\beta$ 4861 or [OIII] 4959, 5007. In most spectra, the H$\beta$ line and the [OIII] 4959, 5007 doublet are usually seen as a triplet of lines (the line ratios of the latter two are fixed by atomic physics at 1:3). Furthermore, given the wide wavelength range of our spectra, either H$\alpha$ or [OII] 3727 should be visible with H$\beta$ for all redshifts (see Figure 1b).

Turning to [OII] 3727, this line can be seen on our spectra over a wide range of redshifts, and could be expected to be the only strong feature present at z > 0.7 when the H$\beta$ 4861 and [OIII] 4959, 5007 combination moves out of the spectral range. It is noticeable and suggestive in Figure 1a that most of the "single emission lines" are indeed seen at $\lambda > 6300$Å corresponding, for an [OII] 3727 identification, to just this range z > 0.7. The Balmer absorption lines that might be expected to be seen near [OII] 3727 are variable in strength and can be filled in by emission.

However, for lines at $\lambda > 6563$Å, a serious potential ambiguity exists between H$\alpha$ at low redshift and [OIII] 3727 at $z > 0.76$. If the line is H$\alpha$, then we might hope to see [SII] 6724 to the red and the H$\beta$ 4861 and [OIII] 4959, 5007 at shorter wavelengths (see Figure 1b) but the absence of the latter can not be taken as definitive due to the possibility of extinction. A final, more exotic, possibility is that such a line is Ly$\alpha$ at $z > 2$. However, for most of the single emission lines, where the line is longward of 6300Å (see Figure 1a), this would make the redshift very high, $z > 4$, and would place the Lyman limit well within our spectral range, at $\lambda > 5000$Å. At these redshifts, extrinsic as well as intrinsic HI absorption would be likely to extinguish the continuum below the Lyman limit and the fact that this is not seen makes this identification unlikely. In addition, if the lines were Ly$\alpha$, it might be expected that more such lines would be seen at $\lambda < 6300$ Å.

In the light of the above discussion, any isolated emission lines at $\lambda < 6560$Å (a minority of the sample - see Figure 1a) were regarded as securely identified with [OII] 3727. In order to secure identifications of the single emission lines at $\lambda > 6560$Å, where there is the potential ambiguity with H$\alpha$, we have investigated whether the shape of the continuum in the neighborhood of the line could be used as a diagnostic. This was motivated by the fact that the continua of many of the single emission line objects show a distinctive rise just redward of the line that resembles that seen in objects where the [OII] 3727 identification is secure on the basis of absorption lines or other emission lines. It is important to note that the wavelength and shape of this continuum feature varies since it is caused by a combination of the general Balmer break in hot stars and the well-known 4000 Å break in cooler stellar populations.

Accordingly, we defined a continuum color and break index as follows. If a single isolated emission line was observed at wavelength $\lambda_{em}$, then a straight line was fit to the continuum (in linear units of $f_\nu$ vs. $\lambda$) between $0.858\lambda_{em}$ and $0.966\lambda_{em}$ and between $1.073\lambda_{em}$ and $1.127\lambda_{em}$. If the line is indeed [OII] 3727 then these ranges correspond to 3200-3600Å and 4000-4200Å. The upper continuum meaurement if therefore above both the Balmer break and the 4000Å break. This procedure is illustrated in Fig 2. We then calculated the flux densities at points A and B (the midpoints of these two ranges) and at point C (the extrapolation of the short wavelength continuum to $1.100\lambda_{em}$, equivalent to 4100 Å), and defined the color and break index to be:

$$\text{color} = -2.5 \log_{10}\left(\frac{f_A}{f_B}\right)$$

$$\text{break} = -2.5 \log_{10}\left(\frac{f_B}{f_C}\right)$$

It should be noted that the "break index" measures the curvature of the continuum and not simply a color across a break (c.f. for example the 4000 break index of Hamilton 1984) and is sensitive to the presence of both the 4000 A break and the Balmer limit at 3648 A. With our spectra, these parameters could be measured for galaxies in which $5600 < \lambda_{em} < 7520$ A (corresponding to $0.5 < z < 1.0$ for an [OII] 3727 identification). Outside of this range, there was inadequate continuum longward or shortward of the line to reliably measure these continuum parameters. We measured these parameters for all the single emission lines in this range and also for all objects in which either [OII] 3727 or H$\alpha$ had been identified to lie in this same wavelength range. These latter form a "defining sample" for our algorithm. We have not attempted to determine error bars for individual measurements due to the fact that the measurements are likely to be affected by non-Poissonian effects. Rather, we take an empirical approach, using the "defining samples" to determine the range of parameters observed in the sample after the inclusion of all observational errors.

### 2.1. Emission line objects which have secure redshifts

Figures 3(a) and 3(b) show these parameters for two defining samples. First, in Figure 3(a) we plot those galaxies with $0.75 < z < 1.0$ which have secure [OII] 3727 line identifications (based on other features) in the $6560 < \lambda_{em} < 7520$ Å range. The measurements in this sample thus span exactly the same wavelength range as in the target sample and thus will reflect to precisely the same degree any problems encountered in defining the continua in the spectra. However, this defining sample may, by definition, be missing prototypes of the objects in which only a single line can be seen. Therefore, in Figure 3b, we plot all emission line galaxies (including the single emission line objects) lying between $0.5 < z < 0.75$, i.e. with $5600 < \lambda_{em} < 6560$ (where the line cannot be H$\alpha$). *These galaxies must span the full range of the population.* The greater scatter in this sample is probably due to the difficulties of measuring the continuum slope at short wavelengths in our spectra. Finally, we plot on both diagrams low redshift galaxies in which H$\alpha$ was observed at $\lambda < 7520$ Å.

On both diagrams there is a clear separation between the high redshift [OII] 3727 emitters and the low redshift H$\alpha$ emitters - the latter do not show a break feature and are also generally bluer in this color. The lines shown on Figure 3 define regions where, empirically, the lines are exclusively identified with [OII] 3727 and H$\alpha$ respectively plus an intermediate, ambiguous, region.

### 2.2. Model spectra

Theoretical models support this empirical analysis. In Figure 3(c) we plot the same indices calculated for a wide range of spectral energy distributions of model galaxies produced by the GISSEL spectral synthesis code of Bruzual and Charlot (1993) with $\lambda_{em}$ set to 3727 Å, 5007 Å, and 6563 Å, respectively (we are greatly indebted to Gabriela Mallen-Ornelas for performing these model computations). In Figure 3(c) we show the color and break indices for the models for the three different $\lambda_{em}$ values. As expected, although there is degeneracy for the bluest models (very young galaxies have blue featureless spectra over a wide wavelength range), the continua around 3727 rapidly develop a significant break and separate from the 5007 and 6563 models in the figure, supporting the empirical evidence from Figures 3(a) and 3(b).

### 2.3. Application to single emission line objects

Finally in Figure 3(d), we plot the indices for the 35 single emission line objects with $6560 < \lambda_{em} < 7520$ Å, i.e. the target sample for which we hope to use the continuum shape to secure the line identification. Apart from five galaxies which lie in the overlap region, these objects lie in the area in which the emission line is identified with [OII] 3727. Based on this, the 30 galaxies in the [OII] area were assigned to a confidence class of "8" (see CFRS II) and we treat them as securely identified at the redshift obtained by identifying the line as [OII] 3727. The other five were assigned a confidence class of "9" and provisionally also placed at the higher redshifts, although the lower H$\alpha$-based redshift cannot be ruled out.

We were unable to use the continuum shape in this way for (a) single emission line objects in which the emission line lay at $\lambda_{em} > 7520$ Å, where we have inadequate long wavelength continuum to reliably determine the break and color indices and (b) the handful of MARLIN spectra with $6563 < \lambda_{em} < 7520$ where the continua were less reliable due to poorer flux calibration. We have assigned these 24 objects also to class "9".

However, the fact that in those objects for which we *could apply* this continuum-based criterion, the line was to a very large degree identified with [OII] 3727 (85% secure identifications and the remainder ambiguous) suggests that *most* of the emission lines in the remaining 29 class "9" objects, where we could not apply this test or where the test was ambiguous, are also likely to be associated with [OII] 3727, and we will generally make this assumption in future papers.

The high degree of association between single isolated emission lines and [OII] 3727 is consistent with the idea that H$\alpha$ should be accompanied by other emission lines (e.g. H$\beta$ and [OIII] 4959, 5007) or absorption lines (e.g. Mg b 5175) leading to a secure identification.

## 3. Analysis of repeat observations

During the course of this project, 187 objects were observed spectroscopically on more than one occasion. These repeat observations were made (a) because the first spectrum had not resulted in an identification, or (b) in order to verify identifications of low confidence class or, in a few cases, (c) by mistake. Regardless of the motivation, the new spectra were reduced and identified in the manner described in CFRS II without knowledge of what the earlier identification had been or even, in most cases, whether the observation was even a repeat. After the analysis of the new spectrum was finalized, the two (or more) spectra of each object were then compared, co-added if this was deemed advantageous, and a final identification and confidence class assigned. In some cases the co-added spectra reinforced the original identifications and in these cases the confidence class was sometimes increased. In a few cases, the coadded spectra did not support the original identifications, particularly if the original identifications were discrepant, and in this case the confidence class could be decreased and/or the object regarded as unidentified.

The fact that each observation was inititially treated quite independently of any earlier ones means that these repeat observations allow a number of empirical tests of the data to be made, and these are described below.

### 3.1. Empirical calibration of the reliability of the confidence classes

Of the 187 objects observed more than once, 91 were independently identified (with a confidence class of 1, 2, 3, 4, 8 or 9 or the quasar classes 12, 13, 14 - see CFRS II) more than once. These objects thus provide an empirical calibration of the reliability of identifications in these confidence classes.

This analysis was carried out as follows: For each individual independent identification in turn, we asked whether this identification was either confirmed or refuted by another identification with equal or higher confidence class, defining this to be a "test". We thus disregarded all identifications made at lower confidence as not significantly testing the higher confidence identification. Where both observations had the same confidence class we regarded this as two "tests". In the case where the two identifications agreed, we counted this as two "successes". If they disagreed, then we compared each with the final identification obtained from coadding both spectra (regardless of its confidence class). This sometimes led to one "success" and one "failure" but sometimes to two "failures" if the coadded spectra supported neither identification and the object had been finally classified as unidentified. As a final complication, the "single emission line" classes 8 and 9 (see Section 2) were regarded as both confirming, and being confirmed by, confidence classes 2, 3 and 4. Under these criteria, we had 147 separate "tests".

The results of these tests are shown in Table 1. The verification rates for the main classes increase monotonically from 56% for class 1 (not regarded as a secure identification) up to 100% for class 4. The verification rates for classes 2 and 3 (81% and 97% respectively) may in fact be underestimates, since 4 of the 6 failures in the class 2 category come from just two objects (in which neither of the two original class 2 identifications were supported by the co- added spectra), and the single failure in class 3 is a rather pathological object. It is one of the few quasars in the sample and the Mg 2799 line at z = 1.35 was initially mistakenly identified in an earlier spectrum as [OIII] 5007 at lower redshift. Finally it should be noted that the emission lines in the single emission line objects were *invariably* confirmed in repeated observations, indicating that our discrimination against cosmic rays was successful. In some cases, the redshift was secured by the identification of other supporting features.

As described in CFRS II, the confidence class scheme was set up at the start of the project to describe, as well as possible, the likelihood that each claimed identification was correct. Reassuringly, the empirical verification rates found in this analysis reflect our original *a priori* definition of these probabilities.

### 3.2. Redshift accuracy

A total of 48 galaxies were identified at confidence note 2 or higher from more than one observation. The differences in the measured redshifts from these independent spectra therefore gives an empirical estimate of the accuracy of the redshifts in the survey. The histogram of redshift differences is shown in Figure 4. There is a broad peak out to $\Delta z = 0.002$ with only a few outliers beyond. The r.m.s. redshift difference is nominally 0.0026 implying a redshift r.m.s. uncertainty per observation of 0.0019 or a velocity uncertainty of 570 kms$^{-1}$. The true uncertainty is likely to be smaller. Elimination of the outliers with $\Delta z > 0.005$ yields an r.m.s. difference of 0.00185, or an r.m.s. uncertainty of 0.0013. An uncertainty of 0.0013 is also obtained by considering the differences in the redshifts of only the highest quality spectra (confidence class 4). This is equivalent to a velocity uncertainty of 390 kms$^{-1}$.

### 3.3. Recovery of initial failures

A final statistic of interest is how often a repeat observation of an initial "failure" (i.e. an object for which no identification could be made from a single 8 hour spectrum) led to it's eventual secure identification, either from the new spectrum alone or through the coaddition of the two spectra. Accordingly, we examined the *final* identification status of the 99 objects with repeated observations in which one observation had failed to securely identify the object (i.e. with confidence class 0 or 1). In 70 of these 99 objects (70%) the additional data resulted in a secure identification in the final catalogue (i.e. with confidence class 2 or greater).

The significance of this statistic is that it means that many of the objects in the final catalogue which were not identified, and which were not reobserved, could presumably have been successfully identified with repeat observations. Furthermore, if it is assumed that the failures which were reobserved are representative of the failures which were not reobserved, then we can use the identifications of the objects "recovered" by the repeat observations to estimate, at least in a statistical sense, the nature of 70% of the failures which were not reobserved. This aspect is developed in CFRS V.

### 4. Spectroscopic data in the 1415+52 and 2215+00 survey fields

Tables 2 and 3 present spectroscopic identifications for objects in the 1415+52 and 2215+00 survey fields. The columns in Tables 2 and 3 are (1) identification number, (2) right ascension [2000.0], (3) declination [2000.0], (4) isophotal $I_{AB}$ magnitude, (5) $(V-I)_{AB}$ color, (6) the Q compactness parameter (see CFRS I), (7) the redshift, with 9.999 indicating an unidentified object and 0.000 indicating a star, (8) the confidence class (see CFRS II) and (9) the spectroscopic features identified. A "1" indicates the general continuum shape supported the identification. A "2" represents the features of an M-star. Figures 5 and 6 show greyscale representations of the I-band images of these fields and Figures 7 and 8 identify those objects listed in Tables 2 and 3.

In the 1415+52 field there are 224 objects in the statistically complete sample plus an additional 14 objects in the supplementary catalogue (see CFRS II for a definition of the latter). The photometric catalogue in the 14hr field contains 560 objects with $17.5 < I_{AB} < 22.5$ so the sampling rate is approximately 40%. The 14hr field benefitted from a prolonged sequence of observations allowing reobservations of unidentified objects. Of the 224 objects in the statistically complete sample, only 27 remain unidentified (i.e. classes 0 and 1) and the spectroscopic identification rate in this field is thus 88% (the highest in the CFRS survey). The fraction of stars is 21%. We have been able to identify many of the faint galaxies in this field with faint $\mu$Jy radio sources from Fomalont et al (1991) and these are discussed in Hammer et al (1995b, CFRS VII). Le Fèvre et al (1994) have described a structure of over a dozen galaxies at z = 0.985 in this field. The overall redshift distribution in this field is shown in Figure 9.

As discussed in CFRS V, examination of the colors and morphologies of two objects in the 1415+52 field, 14.0664 and 14.0823 indicate that they were misclassified spectroscopically as early type galaxies at z $\sim$ 0.3 when in fact they are almost certainly K-stars. Interestingly, the spectra of these two objects, and only these two, had been flagged as suspicious during a final review of all 1010 CFRS spectra (by SJL) after the completion of the three-fold comparisons of independent reductions described in CFRS II. Since only one other object (in the 0000+00 field) in the 1010 spectra in the CFRS appears to have been misclassified in this way, we have simply altered the entries in Table 2 for these two objects.

In the 2215+00 field, there are 189 objects in the statistically complete sample plus an additional 10 objects in the supplementary catalogue. The photometric catalogue in the area surveyed spectroscopically contains 731 objects, so the statistically complete sample represents a sampling rate of 26%. There are 29 spectroscopically unidentified objects and the spectroscopic identification rate is thus 85%. The stellar fraction in this relatively low latitude field is 35% and this largely accounts for the 30% higher surface density of objects in the photometric catalogue of this field relative to the 1415+52 field described above. The redshift distribution in this field is shown in Figure 10.

The 2215+00 field contains the SSA-22 field described by Lilly et al (1991). It and the surrounding areas have also been extensively studied by the Hawaii group, allowing some external checks on our spectroscopic identifications.

## 5. Comparison with spectroscopic identifications by the Hawaii group

Unfortunately, excluding objects previously published by Lilly et al (1993) and Lilly (1993), there are only 8 objects in common between the CFRS sample and that of Songaila et al (1995). As shown in Table 4, the identifications agree in 5 cases with a mean and r.m.s. redshift difference of 0.0002 and 0.0023 respectively. The r.m.s. is similar to that found above. For two of the sources one or other group did not identify the object. For the eighth source, although the identifications

– 9 –

disagree, they are both insecure identifications (confidence class 1 in the CFRS, a class that we have not regarded as a "usable" redshift determination, and ":" in Songaila et al 1995).

## 6. Conclusions

We have investigated whether the curvature of the continuum in the neighborhood of the emission line can be used to identify the emission line in objects in which a single isolated emission line is the distinguishing spectroscopic feature, especially for objects with $\lambda_{em} > 6560$ A where there is potential ambiguity between [OII] 3727 at z $> 0.76$ and H$\alpha$ at low redshift.. We find a clear separation on a two parameter plot between objects in which the identification of the line is secure on the basis of other features. Applying this method to the 35 single emission line galaxies with $6560 < \lambda < 7520$, we find that in at least 85% (and up to 100%) of these galaxies the line should be identified with [OII] 3727. We argue that this most likely applies also for galaxies in which the line is at $\lambda_{em} > 7520$ Åwhere we are unable to adequately sample the continuum.

We have also studied the statistics of the independent identifications of objects that were observed on more than one occasion. The independent identifications of objects that were identified more than once can be used to test or empirically calibrate the confidence classes that we have used in the CFRS to describe the reliability of the spectroscopic identifications. We find good agreement between the success rates and the initial definition of each confidence class. In addition, we find that repeated observations are frequently ( 70%) able to secure an identification for an object which was initially unidentified after one observation. This implies that many of the objects in the survey that have not been identified could be identified if they were reobserved.

We thank Gabriela Mallen-Ornelas for her assistance with the computations of the GISSEL spectra. We also thank the Directors of CFHT and our respective national TACs for their support of the CFRS project, and the referee for his careful reading of the text. SJL's research is supported by NSERC of Canada and the project has been facilitated by a travel grant from NATO.

Table 1: Empirical Validation of Confidence Classes

| Confidence Class | % of complete sample | Number of "tests" | Number of successes | Success rate |
|---|---|---|---|---|
| 4 | 35 | 23 | 23 | 100% |
| 8 and 9 | 6 | 20 | 20 | 100% |
| 3 | 33 | 38 | 37[a] | 97% |
| 2 | 10 | 32 | 26[b] | 81% |
| 1 | 6 | 34 | 19 | 56% |

[a]Single failure was a quasar originally misidentified as a low redshift galaxy.

[b]Four of these six failures are due to two objects that failed twice (see text).

Table 2: Spectroscopic catalog in the 1415+52 field

| Name | RA | DEC | $I_{AB}$ | $(V-I)_{AB}$ | Q | z | Class | Features |
|---|---|---|---|---|---|---|---|---|
| 14.0020 | 14 18 23.71 | +52 30 09.7 | 21.09 | 1.76 | 0.87 | 0.0000 | 4 | 2 |
| 14.0024 | 14 18 23.64 | +52 29 56.9 | 20.67 | 1.11 | 4.36 | 0.5313 | 4 | 3727 3933 3969 4102 |
| ...... | | | | | | | | |

Table 3: Spectroscopic catalog in the 2215+00 field

| Name | RA | DEC | $I_{AB}$ | $(V-I)_{AB}$ | Q | z | Class | Features |
|---|---|---|---|---|---|---|---|---|

Table 4: Comparison of Spectroscopy with Songaila et al. in the 2215+00 field

| CFRS | | | HDS | |
|---|---|---|---|---|
| Name | z | Class | ID | z |
| 22.1210 | 0.417 | 1 | 265 | 0.304: |
| 22.1228 | 0.415 | 2 | 239 | ... |
| 22.1280 | 0.351 | 2 | 231 | 0.348 |
| 22.1294 | ... | 0 | 251 | 0.822 |
| 22.1339 | 0.385 | 4 | 280 | 0.384 |
| 22.1350 | 0.511 | 4 | 283 | 0.513 |
| 22.1417 | 1.102 | 9 | 259 | 1.010: |
| 22.1433 | 0.300 | 2 | 273 | 0.303 |

– 13 –

Fig. 1.— Upper panel. The distribution of wavelengths of the emission lines in spectra in which the only sharp feature was a single isolated emission line. The vertical bars indicate the range of wavelengths over which the analysis of the continuum shape described in the text could be performed. Lower panel. The wavelengths of the most prominent emission lines in the optical spectra of galaxies as a function of redshift. Vertical dashed lines show the spectral range of our spectra. These curves show (a) that [OII] 3727 is likely to be the only emission feature visible for $0.7 < z < 1.3$ and (b) that H$\alpha$ 6563 could be confused with [OII] 3727 for $\lambda > 6563$Å if [SII] 6724 and the H$\beta$, [OIII] 4959, 5007 lines are absent. The continuum analysis was designed to distinguish between these two possibilities.

Fig. 2.— Example of a spectrum showing only a single strong emission line, indicating the continuum measurements made to derive the color and break parameters. See text for details.

Fig. 3.— The break and color indices of the continuum around an emission line, as defined in the text, are plotted for three samples of galaxies and for a set of synthetic galaxy spectra from the GISSEL library. Panel (a) shows galaxies with secure $0.76 < z < 1.03$ in which the emission line is securely identified with [OII] 3727 on the basis of other spectral features (solid points) and galaxies with secure $0.0 < z < 0.15$ in which the line is securely identified with H$\alpha$ (open circles). Panel (b) is the same as (a) except that the high redshift sample now consists of all galaxies with $5590 < \lambda_{em} < 6560$, i.e. $0.5 < z < 0.76$, in which the emission line cannot be H$\alpha$. The increased scatter is due to the difficulty of measuring the continuum at the shorter wavelengths. These two panels both indicate a separation between these indices for spectra in which the line is [OII] 3727 and those in which the line is H$\alpha$ with only a small region of ambiguity (between the sloping lines). Panel (c) shows the indices for a wide range of synthetic spectra generated by the GISSEL package assuming that an emission line is [0II]3727 (open circles), [OIII]5007 (crosses) and H$\alpha$ (small dots) respectively. Except for some degeneracy for very blue spectra (i.e. the youngest model galaxies), the three possibilities occupy separate loci on the diagram. Panel (d) shows the parameters of the test sample consisting of galaxies with only a single emission line. The vast majority of these are in the area occupied by the [OII] 3727 identifications with only a few in the ambiguous area. The scatter in this plot would be expected to be comparable to that in panel (a). See text for further details and discussion.

Fig. 4.— Histogram of redshift differences $|\Delta z|$ derived from multiple observations of galaxies.

Fig. 5.— Greyscale representation of the 1415+52 field. North is at top, East is to the left. Field is 10x10 arcmin$^2$.

Fig. 6.— Greyscale representation of the 2215+00 field. North is at top, East is to the left. Field is 10x10 arcmin$^2$.

– 14 –

Fig. 7.— The 1415+52 field showing objects observed spectroscopically (solid symbols) with other objects with $I_{AB} < 22.5$ marked as open symbols. The symbol size represents the isophotal magnitudes of the objects. The positions are relative to the field center $14^h\ 17^m\ 53\overset{s}{.}73$, $+52°30'31''$. North is at the top, East is to the left.

Fig. 8.— The 2215+00 field showing objects observed spectroscopically (solid symbols) with other objects with $I_{AB} < 22.5$ marked as open symbols. The symbol size represents the isophotal magnitudes of the objects. The positions are relative to the field center $22^h\ 17^m\ 48\overset{s}{.}0$, $+00°17'34''$. North is at the top, East is to the left.

Fig. 9.— Redshift distribution for galaxies in the 1415+52 field. Boxes represent relative numbers of stars and unidentified objects (Confidence Class 0 or 1).

Fig. 10.— Redshift distribution for galaxies in the 2215+00 field. Boxes represent relative numbers of stars and unidentified objects (Confidence Class 0 or 1).

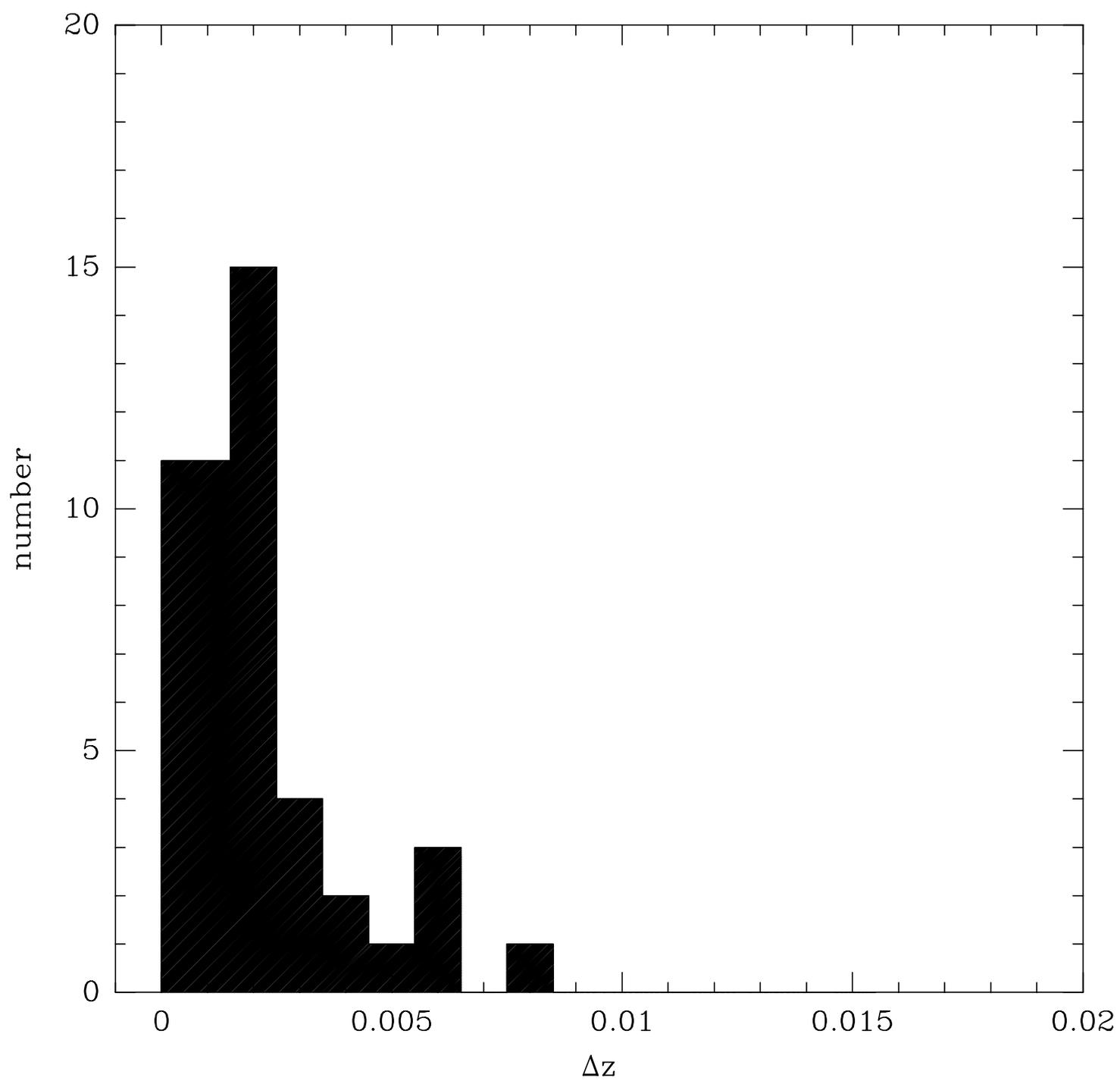

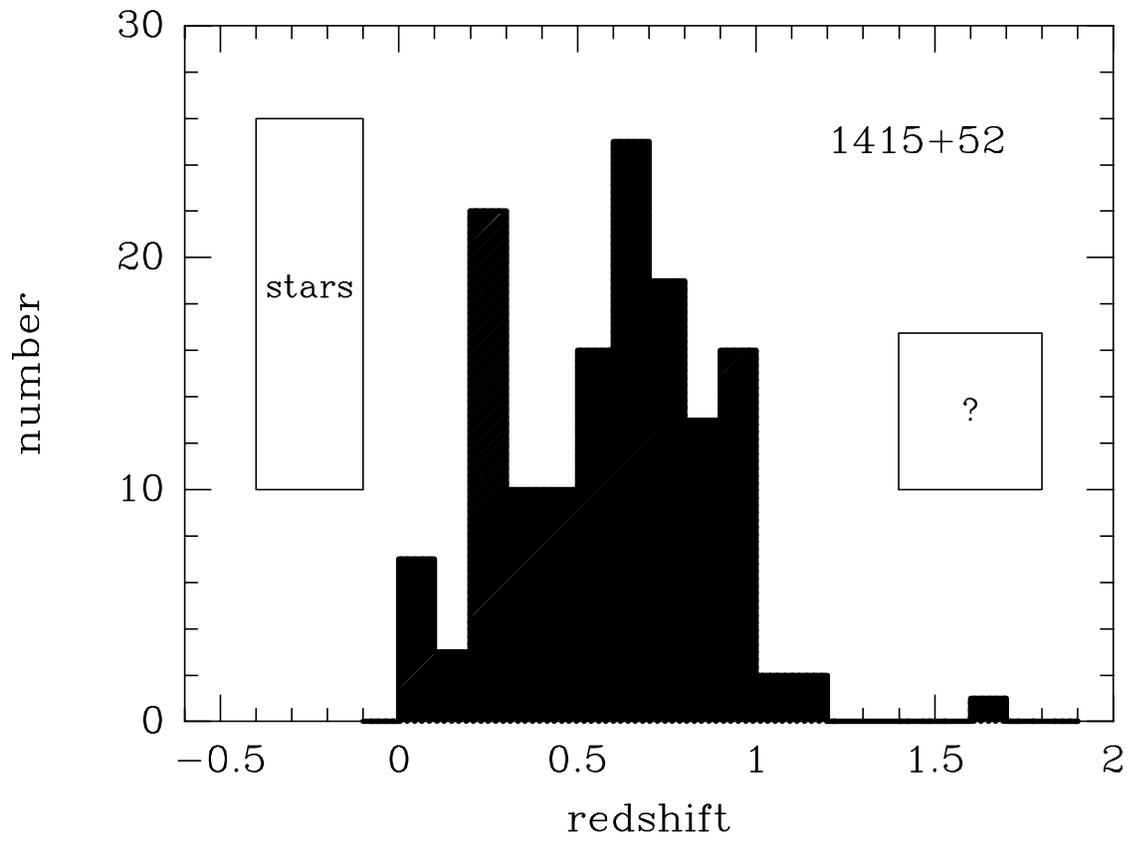

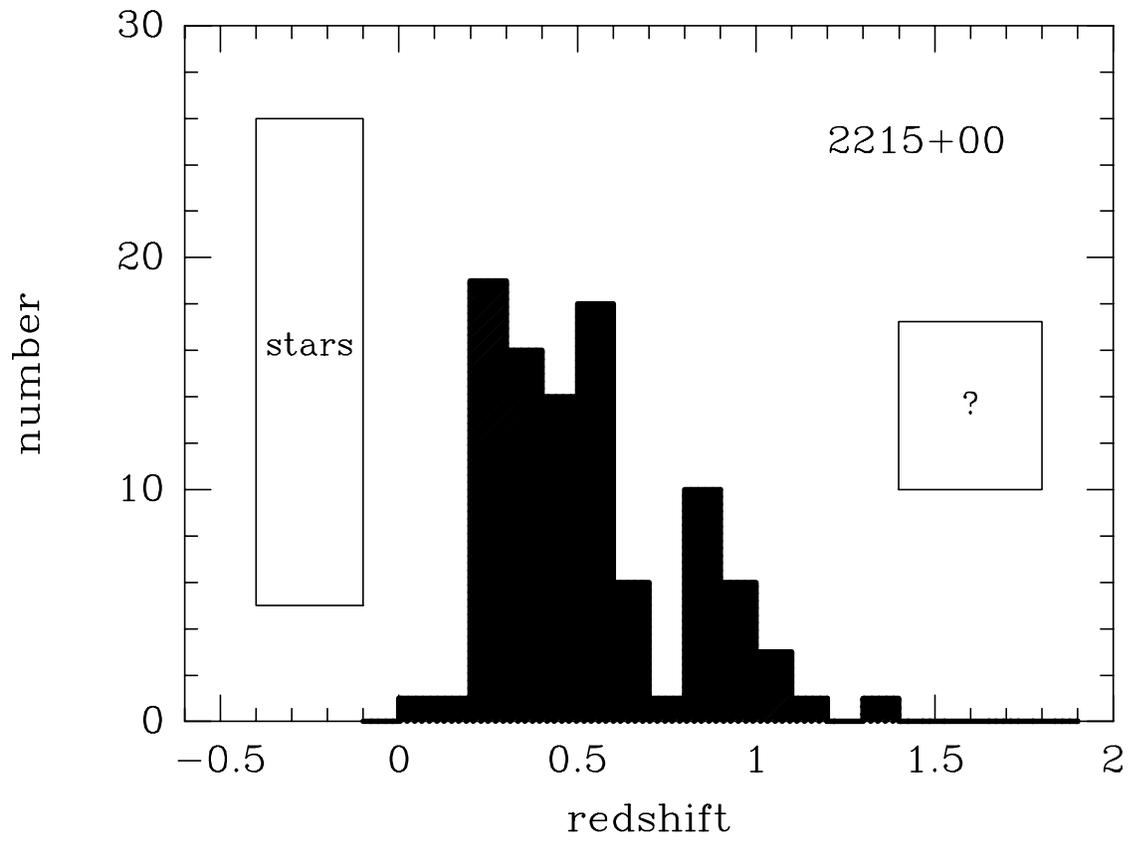